\documentclass[prl,twocolumn,preprintnumbers,nofootinbib]{revtex4-1}%

\usepackage[english]{babel}

\usepackage{amsmath,amsfonts,mathtools,mathrsfs,amssymb}
\usepackage{mdframed}

\usepackage{color}
\usepackage{enumitem}
\usepackage[hidelinks]{hyperref}

\usepackage{tikz}
\usetikzlibrary{trees}
\usetikzlibrary{snakes}
\usetikzlibrary{arrows}
\usepackage{color}

\def\ukap{\underline{\kappa}}
\def\unu{\underline{\nu}}

\newcommand{\beq}{\begin{equation}}
\newcommand{\eeq}{\end{equation}}
\newcommand{\bea}{\begin{eqnarray}}
\newcommand{\eea}{\end{eqnarray}}

\begin{document}

\title{Two-particle spectrum of tensor multiplets coupled to $AdS_3\times S^3$ gravity}

\author{Francesco Aprile}
\email{francesco.aprile1@unimib.it}
\affiliation{{ Dipartimento di Fisica, Universit\`a di Milano-Bicocca \& INFN, 
		Sezione di Milano-Bicocca, I-20126 Milano
		}}
\author{Michele Santagata}
\email{M.Santagata@soton.ac.uk}
\affiliation{School of Physics and Astronomy, University of Southampton, Highfield SO17 1BJ, UK}

\begin{abstract}
\noindent  
We study certain infinite families of two-particle operators
exchanged in 4pt correlators $\langle{\cal O}_{p_1}{\cal O}_{p_2}{\cal O}_{p_3}{\cal O}_{p_4}\rangle$ of tensor multiplets living on the  $AdS_3\times S^3$ background. 
This is the weakly curved, weakly coupled SUGRA theory dual to the D1-D5 system with RR flux. 
At tree level in Mellin space, all these correlators are nicely  determined by a single amplitude, which makes manifest the large 
$p$ limit, the connection with the flat space S-matrix, and a six dimensional conformal symmetry.
We compute the $(1,1)\times \overline{(1,1)}$ superconformal blocks  for the two-dimensional 
$\mathcal{N}=(4,4)$  conformal theory at the boundary, and then we obtain a formula for the 
anomalous dimensions  of the two-particle operators exchanged in the symmetric and anti-symmetric flavor channels. 
These anomalous dimensions solve a mixing problem which is analogous to the one in 
$AdS_5\times S^5$ with interesting modifications. Along the way we show how the 
$(1,1)\times \overline{(1,1)}$ superconformal blocks relate to those in $\mathcal{N}=4$ SYM in four 
dimensions, and provide new intuition on the known data for $AdS_5\times S^5$. 
\end{abstract}
\maketitle


{\bf Introduction.} Understanding what are the possible UV completions of classical 
gravity is one of the most exciting and challenging problems of modern theoretical physics. 
Valuable help might come from solving the same problem,  but in spaces 
with an $AdS$ factor, where the $AdS/CFT$ correspondence plays an important 
role \cite{Maldacena:1997re}. In such circumstances, would a low energy field theorist 
be able to reconstruct the underlying curved string theory, let's say, out of scattering data of gravitons and single particles operators? and how?

Recent work, based on analytic bootstrap techniques in the dual CFT, 
has shown how to address this question in
$AdS_5\times S^5$ SUGRA, away from 
the classical regime  \cite{Rastelli:2016nze,Aprile:2017bgs,Alday:2017xua,
Aprile:2017qoy,Alday:2018pdi, Drummond:2019hel,Aprile:2019rep,Bissi:2020woe,Drummond:2020uni}
and up to one-loop in Newton's constant. 
Most notably, the simplest \emph{one-loop} amplitude for four 
gravitons (supermultiplets) was computed in  \cite{Aprile:2017bgs} by implementing a bootstrap program which, in order to determine the amplitude, used the self-consistently of the operator 
product expansion (OPE) for all spins. The main observation in  \cite{Aprile:2017bgs} was to notice that when scattering states are single particle operators, such as gravitons and Kaluza-Klein modes, 
the operators flowing in the OPE with leading order CFT data are two-particle operators, and the latter can be studied by solving a well defined mixing problem at tree level  \cite{Aprile:2017xsp}.  
The CFT data of the two-particle operator can then be used to construct the leading discontinuities of the one-loop correlator and, with the help of crossing symmetry, bootstrap the full correlator. 

Even though the two-particle bootstrap does not completely fix the amplitude, the left over ambiguities 
were shown to be very constrained, with finite spin support.\footnote{In the case of  four 
gravitons (supermultiplets), the only spin zero ambiguity was resolved
independently in \cite{Binder:2019jwn,Chester:2019pvm}.}
Interpreted as an effective field 
theory computation, this result 
encourages the idea that the presence of
ambiguities, at least in some theories of gravity, might not be as severe as 
naive considerations would suggest.
In fact, the same two-particle bootstrap program was then extended to compute
one-loop 4pt amplitudes of arbitrary external single particle operators, carrying Kaluza-Klein charge under the sphere. The one-loop 4pt amplitudes  so constructed were shown to  
pass spectacular consistency checks \cite{Aprile:2019rep}.

%
With analogous surprise, the spectrum of anomalous dimensions of two-particle operators in $AdS_5\times S^5$ is not completely 
lifted by tree level supergravity \cite{Aprile:2017xsp,Aprile:2018efk}, but remains partially degenerate.
Only $\alpha'$ corrections lift this partial degeneracy uniquely \cite{Drummond:2019odu,Drummond:2020dwr,Aprile:2020mus}, 
despite the fact that the curved Virasoro-Shapiro amplitude \cite{Abl:2020dbx} is not uniquely 
determined within the bootstrap approach, because of ambiguities similar to those mentioned previously. Again, we seem to converge on the concrete possibility 
that a rich mathematical structure lies within $AdS_{d+1}\times S^{d+1}$ supergravity, 
as for the beautiful hidden conformal symmetry discovered in \cite{Caron-Huot:2018kta}. 

In this paper we will continue exploring the structure of $AdS_{d+1}\times S^{d+1}$ gravity, by studying the case of $AdS_3\times S^3$. In particular, by studying tree level amplitudes of (chiral primaries) single particle fields 
in the weakly curved SUGRA regime of the D1-D5 system with Ramond-Ramond fluxes \cite{deBoer:1998kjm,Berkovits:1999im,Taylor:2007hs}. 
Currently, this is the only other SUGRA background in which the spectrum of the dual (strongly coupled) CFT theory can be investigated in great detail with our method. 
In fact, we will proceed in parallel with $AdS_5\times S^5$, first by developing the necessary superconformal block technology,  
with $\mathcal{N}=(4,4)$ superconformal symmetry, and then use it to extract OPE data from 
the tree level correlators $\langle {\cal O}_{p_1}{\cal O}_{p_2}{\cal O}_{p_3} {\cal O}_{p_4}\rangle$ 
bootstrapped in \cite{Giusto:2018ovt,Giusto:2019pxc,Giusto:2020neo,Giusto:2020mup,Rastelli:2019gtj}. 

The single particle operators ${\cal O}_{p}$ living on $AdS_3\times S^3$ originate from  a number $n$ of 
tensor multiplets in the 6d supergravity compactified on $S^3$, 
denoted by $s^I$ in  \cite{Romans:1986er,Deger:1998nm,Mihailescu:1999cj,Arutyunov:2000by}, and therefore have also a flavor index, besides the Kaluza-Klein charge $p$.
%
%
%
Following the unmixing approach of \cite{Aprile:2017xsp,Aprile:2018efk}, we will compute tree level anomalous dimensions 
of certain \emph{two-particle operators with flavor}, denoted afterwards by $\mathcal{O}_{(rs)}^{+}$, exchanged in $\langle {\cal O}_{p_1}{\cal O}_{p_2}{\cal O}_{p_3} {\cal O}_{p_4}\rangle$. 
We will find a very simple answer, which, apart for the factor of $\mathbb{D}$ explained in due course, takes the following form,
\beq\label{intro_anom_dim}
\begin{array}{rl}
{\eta}^{+}_{\tau,l,[ab]}(rs)&=
\displaystyle
\ -\frac{2}{N}\,
%
%
\frac{ \mathbb{D}_{\tau,l,[ab]} }{\rule{0pt}{.35cm}\ \left(\ell_{6d}+1\right)_2}  \\[.2cm] 
\ell_{6d}&=\ l+2r-a-1-\tfrac{1+(-1)^{a+l}}{2}
\end{array}
\eeq
where the the $AdS_3$ quantum numbers $\tau,l$ are the free theory dimension and spin, then $[ab]$ are the $S^3$ quantum numbers, 
and finally $(rs)$ is a pair of integers indexing the two-particle operator. 
Remarkably, $\ell_{6d}$ is essentially the only quantity controlling
the anomalous dimensions, and it has the interpretation of a 6d effective spin. In particular, it only depends on $r$, rather than $(rs)$. 
Thus, similarly to what happens in $AdS_5\times S^5$, 
anomalous dimension with a structure like \eqref{intro_anom_dim}
are degenerate as long as the $AdS_3\times S^3$ quantum numbers give the same value for $\ell_{6d}$. 

The tensor multiplet correlators $\langle {\cal O}_{p_1}{\cal O}_{p_2}{\cal O}_{p_3} {\cal O}_{p_4}\rangle$ on 
$AdS_3\times S^3$ in fact benefit from a hidden 6d conformal symmetry, which nicely resum them all into a single Mellin amplitude, and simply explains the residual degeneracy of the two-particle spectrum in \eqref{intro_anom_dim}.
In $AdS_5\times S^5$ this is a 10d conformal symmetry \cite{Caron-Huot:2018kta}.  
What is intriguing about this parallelism is the way the hidden conformal symmetry actually goes across dimensions.  
We will show indeed that, as a byproduct of our studies here, superconformal blocks for both $AdS_5\times S^5$ and $AdS_3\times S^3$ can be treated 
at once by using the $(1,1)\times \overline{(1,1)}$ formalism that we will introduce.\\


{\bf Tree level correlators.} The correlators we are interested in have a (generalised) disconnected free part and a
\emph{dynamical} contribution of the form
\beq\label{genera_ampl}
\langle {\cal O}_{p_1} {\cal O}_{p_2} {\cal O}_{p_3} {\cal O}_{p_4} \rangle_{dyn.}={\tt kinematics}\times 
\mathcal{A}_{\vec{p}}
\eeq
where $\mathcal{A}$ will denote the \emph{amplitude} of the correlator. 
We will clarify in the next section what {\tt kinematics} stands for, and what are \emph{all} the
allowed superconformal structures, such that $\mathcal{A}$ is function of the cross ratios
$U,V$ in spacetime, and $\tilde U,\tilde V$ on the sphere \cite{cross_ratios}.  
Eq.~\eqref{genera_ampl} is a non perturbative statement, and the $AdS_3\times S^3$ amplitude we will study fits 
\eqref{genera_ampl} for a specific choice of  {\tt kinematics}, hereafter denoted by $\mathcal{A}^+_{\vec{p}}$. 

Parametrising spacetime insertion points with $4d$ embedding coordinates, 
these are given by
\beq
\begin{array}{c}
U=\frac{ X_{12}X_{34} }{X_{13} X_{24}}\quad;\quad V=\frac{ X_{14}X_{23} }{X_{13} X_{24}}
\end{array}
\eeq
with $X_{ij}=X_i.X_j$ and $X_{ii}=0$ for $i,j=1,2,3,4$. Similarly, $\tilde U$ and $\tilde V$ are defined as above with the replacement $X\rightarrow Y$, 
where $Y_i$ are null and parametrise internal space insertion points. 
Then, the natural language to write $\mathcal{A}^+_{\vec{p}}$ is Mellin space \cite{Penedones:2010ue}, upgraded for $AdS_{d+1}\times S^{d+1}$ backgrounds as done in \cite{Aprile:2020luw}.
Indeed, 
the following very compact representation holds for $\mathcal{A}^+_{\vec{p}}$, 
\beq\label{master_mellin}
\mathcal{A}^+_{\vec{p}}=-\oint dsdt \sum_{\tilde s, \tilde t, \tilde u} U^s V^t \tilde{U}^{\tilde s}{\tilde V}^{\tilde t}\,\Big(
\Gamma_{\otimes}\!\times \mathcal{M}_{\vec{p}}(s,\tilde s, \ldots)\Big)
\eeq
where 
\beq
\Gamma_{\otimes}\!=\frac{ \Gamma[- s]\Gamma[-s+c_s]}{\Gamma[1+\tilde s]\Gamma[1+\tilde s+c_s]}\!
\times \frac{ t{\tt -channel }}{\tilde t{\tt -channel } } \times  \frac{ u{\tt -channel }}{\tilde u{\tt -channel } }
\eeq
with\footnote{That $p_3$ appears singled out is conventional, and will have to do with {\tt kinematics}. 
A trivial change of variables can be used to implement permutation symmetric conventions, but there is no point in doing that since superconformal blocks do not have such a symmetry. 
In fact, in both cases another change of variables to $\{{\bold s},\tilde{\bf s},\ldots\}$ is needed to manifest crossing. See eq.~(2.12) of \cite{Aprile:2020mus}.}
\beq
\!\!\!\begin{array}{c}
s+t+u=-p_3-1\quad;\quad  \tilde s+\tilde t+\tilde u=p_3-1 \\[.2cm]
c_s=\frac{p_1+p_2-p_3-p_4}{2}\ ;\ c_t=\frac{p_1+p_4-p_2-p_3}{2}\ ;\ c_u=\frac{p_2+p_4-p_3-p_1}{2}\\[.2cm]
\end{array} 
\eeq
Here $s,t,u$ and $\tilde s,\tilde t,\tilde u$ are a choice of Mellin variables, and the triplet $c_s,c_t,c_u$ accompanies our choice in charge space. 
Note that the sum \eqref{master_mellin} is restricted to the triangle $\tilde s\ge -min(0,c_s)$, $\tilde t\ge -min(0,c_t)$, $\tilde u\ge -min(0,c_u)$
 due to the $\Gamma$ function in the denominator of $\Gamma_{\otimes}$. It can also be turned into a contour 
integral. 
This is useful since it was shown in \cite{Aprile:2020luw} that upon taking $p_{i}$ large 
the integrals localise on a classical saddle point whose action is determined just by $\Gamma_{\otimes}$. 
The result matches the computation of four geodesics shooting from the boundary and meeting 
in the bulk. Reading off the momenta at the intersection point, it was understood that the combinations of Mellin variables
\beq
{\bf s}=s+\tilde s\quad;\quad {\bf t}=t+\tilde t \quad;\quad {\bf s}+{\bf t}+{\bf u}=-2
\eeq
evaluated at the saddle point, become proportional to actual Mandelstam invariants of a 
flat space scattering process in higher dimensions, where the sphere is decompactified.  This nicely explains that $\lim_{p\rightarrow \infty} \mathcal{M}$ 
is fixed by the flat space S-matrix \cite{Lin:2015dsa,Heydeman:2018dje} and provides a canonical covariantisation of 
$\mathcal{M}_{1111}\rightarrow \mathcal{M}_{\vec{p}}$ onto $AdS_3\times S^3$, yielding the result
\beq\label{all_correlators}
\mathcal{M}_{1111}({\bf s},{\bf t}) 
=\frac{\ \delta^{12}\delta^{34}}{{\bf s}+1} +\frac{\ \delta^{14}\delta^{23}}{{\bf t}+1}+\frac{\ \delta^{13}\delta^{24}}{{\bf u}+1}
\eeq
where the $\delta^{ij}\equiv\delta^{I_i I_j}$ are $n$ dimensional Kronecker 
deltas referred to the flavor indexes $I$ that we have been omitting until now.  Upon assuming the 
existence of a 6d conformal symmetry, \eqref{all_correlators} becomes the exact result, 
\beq
\mathcal{M}_{\vec{p}}=\mathcal{M}_{1111}({\bf s},{\bf t}).
\eeq
In particular, we can read off the Mellin amplitude for generic charges $\vec{p}$ out of the very same $\mathcal{M}_{1111}$. 

The flavor structure of the correlators will be decomposed in channels, e.g.~\cite{Giombi:2017cqn,Cordova:2018uop}. 
Thus, we introduce the singlet, $\mathbb I$, the symmetric, $\mathbb S$, and the antisymmetric 
channel, $\mathbb A$. In the order,
\beq
\!\!\begin{array}{rlcc}
\mathcal{M}_{\vec{p}}^{\mathbb I}&=
\frac{1}{n}\!\left[\frac{1}{ {\bf t}+1}+\frac{1}{ {\bf u}+1} \right]+\frac{1}{{\bf s}+1} \qquad ;\qquad \mathbb I=\delta^{12}\delta^{34} \\[.3cm]
\mathcal{M}_{\vec{p}}^{\mathbb S}&=
\frac{1}{2}\left[ \frac{1}{ {\bf t}+1}+\frac{1}{ {\bf u}+1} \right] \quad; \ \ \mathbb S=\delta^{13}\delta^{24}+\delta^{14}\delta^{23}-\frac{2}{n}\delta^{12}\delta^{34}\\[.3cm]
\mathcal{M}_{\vec{p}}^{\mathbb A}&=
 \frac{1}{2}\left[ \frac{1}{ {\bf t}+1}-\frac{1}{ {\bf u}+1} \right]   \quad; \quad\quad\quad \mathbb A=\delta^{14}\delta^{23}-\delta^{13}\delta^{24}\\
\end{array}
\eeq
We will focus on $\mathcal{M}^{f}_{\vec{p}}$ with $f=\mathbb{S},\mathbb{A}$, since these are closed sectors.
For the singlet channel one should include other correlators in the 6d (2,0) supergravity 
\cite{Rastelli:2019gtj,Giusto:2020neo}, such that all allowed two-particle operators participate.\\


{\bf Superconformal symmetry.} The dual conformal field theory that we are studying has 
$\mathcal{N}=(4,4)$ superconformal symmetry in 2d, and the relevant superconformal 
blocks belong to the product $(1,1)\!\times\!\overline{(1,1)}$, where the notation $(1,1)$ 
refers to superconformal blocks of $SU(1,1|2)$, studied in \cite{Doobary:2015gia}. These  superconformal blocks
are labelled by a Young diagram  $\ukap=[\kappa,1^{\kappa'-1}]$ with at most one row and 
one column, of length $\kappa$ and $\kappa'$ respectively, thus $(1,1)$. 
In fact, Heslop and Doobary wrote  in \cite{Doobary:2015gia} a beautiful determinantal 
formula for the more general $(m,n)$ superconformal blocks. Borrowing that result, 
we introduce the $(1,1)$ superconformal block
\beq\label{formula_F}
\!\!
\begin{tikzpicture}
\draw (0,0) node {$\displaystyle \!B^{(\alpha,\beta)}_{\gamma,\ukap}  =  \underbrace{
  g_{12}^{\frac{p_1+p_2}{2} } \!g_{34}^{\frac{p_3+p_4}{2} } \left[ \frac{ g_{14} }{g_{24} }\right]^{\frac{p_1-p_2}{2}}\!\left[ \frac{ g_{14} }{g_{13} }\right]^{\frac{p_4-p_3}{2}}
\!\!\!\bigg( \frac{x}{y}\bigg)^{\!\!\frac{\gamma}{2}} }  F_{\gamma,\ukap}^{(\alpha,\beta)}$};
\draw (.3,-.9) node[scale=.9] {\tt prefactor$_{\gamma}$};
\end{tikzpicture}
\eeq
where $\alpha=max$ and $\beta=min$ $@$ $( \frac{ \gamma-p_{12} }{2}, \frac{\gamma-p_{43}}{2})$, 
\beq
F_{\gamma,\ukap}^{(\alpha,\beta)}
  \displaystyle  =   \delta_{\ukap,\underline{0}}  \bigg( \frac{y}{x}\bigg)^{\!\!\beta} +  (x-y) H_{\ukap}(x,y) 
\eeq
and the dependence on $\ukap$ enter through
\beq\label{formula_H}
H_{\ukap}=
\left\{\begin{array}{cc}    
				\sum_{k=0}^{\beta-1}\ h^{(\alpha,\beta,\gamma)}_{-k}(x) h^{(-\alpha,-\beta,-\gamma)}_{k+1}(y) & \,\ukap=\underline{0} \\[.4cm]
				(-)^{\kappa'-1} h^{(\alpha,\beta,\gamma)}_{\kappa}(x)h^{(-\alpha,-\beta,-\gamma)}_{\kappa'}(y) &\ {\rm otherwise}
\end{array}\right.
\eeq
with
$h_{k}^{(a,b,c)}(z)=z^{k-1}~_2F_1(k+a,k+b;2k+c;z)$. The parameter $\gamma$ specifies, together with the Young diagram $\underline{\kappa}$, the exchanged representation. 
It plays an important role for short representations. However, since we will be mainly interested in long representation, it will not be essential in our discussion. 
It is nevertheless useful to understand his origin diagrammatically, for example in free theory. 
Indeed, as it appears in the prefactor in \eqref{formula_F}, it counts the powers of cross ratios $x/y=(g_{13}g_{24})/(g_{12} g_{34})$, and therefore  the number of propagators going from points $(12)$ to $(34)$, so we can think of it as setting the origin for the dimension of the exchanged operators in that diagram.

Note that $F_{\gamma,\ukap}$ has a polynomial expansion in both $x$ and $y$.  In particular, 
$F_{\gamma,\varnothing}=1+\ldots$, since the resummation  in $x$ is non trivial whenever 
$\beta\ge1$. Note also that $\kappa'\leq \beta$,  otherwise  $F_{\gamma,\ukap}$ vanishes (in the $y$). 
In fact the $F_{\gamma,\ukap}$ is better defined as an expansion 
over super Schur polynomials of the form
\beq
		F_{\gamma,\ukap}=\sum_{ {\unu}:\ukap\subseteq \unu} (T_{\gamma})_{\ukap}^{\unu}\ s_{\unu}(x|y)
\eeq
where, by construction, $s_{\unu}(x|y)$ and their multivariate 
generalisation, solve the superconformal Ward identity,
\beq
\Big[(\partial_{x_i} +\partial_{y_j})s_{\unu}\Big]_{x_i=y_j}=0
\eeq
and thus the superconformal block so constructed.

A basis for the $\mathcal{N}=(4,4)$ superconformal blocks  is obtained  by taking products 
of such $F$. On the the real slice, we will distinguish among,
\beq\label{table_products}
\ \ \ \begin{array}{l}
\begin{tikzpicture}
\def\step{1}
\draw[] (-1,-1.1) rectangle (5.9,1);
\node at (.5*\step+.5, 0.62) {$B_{\varnothing}(x,y)B_{\varnothing}(\bar x,\bar y)$};
\node at (.5*\step+.5, -0.05) {$B_{\ukap}(x,y)B_{\varnothing}(\bar x,\bar y)+c.c.$};
\node at (.5*\step+.5, -0.75) {$B_{\ukap_1}(x,y)B_{\ukap_2}(\bar x,\bar y)+c.c.$};
\draw[] (-1+4,-1.1)  -- (-1+4,1);
\node at (4.2*\step+.25, 0.65) {half-BPS};
\draw[] (-1,.3) -- (5.9,.3);
\node at (4.2*\step+.25, 0) {short};
\draw[] (-1,-.4) -- (5.9,-.4);
\node at (4.2*\step+.25, -0.7) {long};
\end{tikzpicture}
\end{array}
\eeq
In each of these cases the result always fits into the form
\beq\label{more_gen_G}
\mathcal{G}={\cal C}+\Big[ (x-y) \mathcal{S}(x,y) + c.c.\Big]+ (x-y)(\bar x-\bar y) \mathcal{H}(x,\bar x,y,\bar{y})
\eeq
where ${\cal C}$ is a constant, while $\mathcal{S}$ and $\mathcal{H}$ 
are the \emph{single}- and \emph{two}- variables contributions, respectively.

The result \eqref{more_gen_G} is also quite intuitive. 
The special factors $(x-y)$ and $(\bar x-\bar y)$ are simply
$s_{\Box}(x|y)=(x-y)$  for a single box Young diagram, and its complex conjugate. Moreover, they vanish on $x=y$ or $\bar{x}=\bar{y}$. 
Thus $\mathcal{H}$ is the part of the 
correlator which vanishes when both $x=y$ and $\bar{x}=\bar{y}$. 
The rest, necessarily goes with $(x-y)$ or $(\bar{x}-\bar{y})$, unless it is a constant. 
In this way, it is also simple to see that the $\mathcal{N}=(4,4)$ Ward Identity,
\beq
\Big[(\partial_x+\partial_y) \mathcal{G}\Big]_{x=y}=0\quad;\quad \Big[(\partial_{\bar x}+\partial_{\bar y}) \mathcal{G}\Big]_{{\bar x}={\bar y}}=0
\eeq
is satisfied for any ${\cal C},{\cal S}$ and ${\cal H}$.\\

{\bf Long representations.}
Long superconformal blocks factorise into  their bosonic components, 
i.e.~conformal and internal.  To see this, take \eqref{table_products} and change basis 
by considering linear  combinations of the form
\beq\label{change_b_block}
\tfrac{1}{2}\left( 
B_{[\kappa_1,1^{\kappa'_1-1}]} \overline{B}_{[\kappa_2,1^{\kappa'_2-1}]} \pm {B}_{[\kappa_1,1^{\kappa'_2-1}]}\overline{B}_{[\kappa_2,1^{\kappa'_1-1}]} \right)+c.c.
\eeq
This change of basis leads to the general decomposition
\beq\label{H_split}
\!\!\!\!\mathcal{H}(x,\bar x,y,\bar{y})=\mathcal{H}^+(U,V,\tilde U,\tilde V)+(x-\bar x)(y-\bar y) \mathcal{H}^-(U,V,\tilde U,\tilde V)
\eeq 
where $\mathcal{H}^{\pm}$ will now have a clear relation with bosonic blocks, 
since they are symmetric in $x,\bar x$ and $y,\bar y$, and therefore writable as function of  $U,V$ and $\tilde U,\tilde V$. 
Before giving more details, let us make a first remark: The most general form of a 
$\mathcal{N}=(4,4)$ correlator,  for four half-BPS external particles, is necessarily 
given by $\mathcal{G}$ in \eqref{more_gen_G}, with the splitting of $\mathcal{H}$ 
as in \eqref{H_split}.  The dynamical correlator in \eqref{genera_ampl} 
thus admits two types of kinematics, 
\beq
\begin{array}{c}
{\tt kinematics}^+={\tt prefactor}_{p_3+p_4}\times (x-y)(\bar x-\bar y) \\[.2cm]
{\tt kinematics}^-=(x-\bar x)(y-\bar y)\times {\tt kinematics}^+
\end{array}
\eeq
The tree level correlator $\mathcal{A}_{1111}$ of \cite{Giusto:2018ovt,Giusto:2019pxc,Giusto:2020neo,Rastelli:2019gtj}
has $\mathcal{A}^+_{1111}\neq 0$, and it was shown that $\mathcal{A}^-_{1111}$ is absent, correctly. In fact, there is no room for $\mathcal{A}^-_{}$, as a polynomial in $y,\bar{y}$, with such a minimal assignment of charges. 
Assuming a 6d conformal symmetry, $\mathcal{A}^+_{1111}$ is promoted to a  generating 
function for $\mathcal{A}^+_{\vec{p}}$ through \eqref{all_correlators}, but $\mathcal{A}^-_{\vec{p}}$ 
cannot be generated this way.\\


{\bf On the $\mathcal{N}=4$ superconfomal blocks.} To analyse $\mathcal{A}^+_{\vec{p}}$, we will need an explicit formula for 
$\mathcal{H}^{\rm long +}_{\kappa_1,\kappa_2,\kappa'_1,\kappa'_2}$.  From \eqref{formula_F}-\eqref{formula_H} we find
\beq\label{Hpluslong}
\begin{array}{l}
		\!\!\!\!\mathcal{H}^{\rm long +}_{}=\\
		\displaystyle    
		\bigg( \frac{y\bar y}{x \bar x}\bigg)^{\!\!\frac{\gamma}{2}}\ 
			\frac{ (-)^{\kappa'_1-\kappa'_2}\ {\tt B}^{(+p_{12},+p_{43})}_{\kappa'_1-\frac{\gamma}{2},\,\kappa'_2-\frac{\gamma}{2}, }(y,\bar y) }{ \rule{0pt}{.35cm} \tilde U^{} }\ 
			\frac{{\tt B}^{(-p_{12},-p_{43})}_{\kappa_1+\frac{\gamma}{2},\, \kappa_2+\frac{\gamma}{2} }(x,\bar x )}{ \rule{0pt}{.35cm} U^{}} 
\end{array}
\eeq
with the bosonic (and normalised) block \cite{Dolan:2003hv}
\beq
{\tt B}^{(a,b)}_{k_1k_2}(z,\bar z) =
 \frac{ z  h_{k_1}^{ (\frac{a}{2},\frac{b}{2},0) }(z)\, {\bar z} h_{k_2}^{ (\frac{a}{2},\frac{b}{2},0) }(\bar z)  + c.c. }{2(1+{\delta_{k_1k_2} })}
\eeq
Note that {\tt prefactor}$_\gamma\times$\eqref{Hpluslong} does not depend on $\gamma$ anymore,
since it can be absorbed into the $SO(2,2)\times SO(4)$ quantum numbers of $\mathcal{H}^{\rm long +}$, which are
\beq\label{change_qnumbers}
\begin{array}{ccc}
1+\frac{\tau}{2}=\frac{\gamma}{2}+\kappa_2&\quad;\quad & l=\kappa_1-\kappa_2\ge 0   \\[.2cm]
\ \ \ \ \ \frac{b}{2}=\frac{\gamma}{2}-\kappa'_1 &\quad;\quad & a=\kappa'_1-\kappa'_2\ge 0
\end{array}
\eeq
where recall that $\kappa_{i=1,2}\ge 1$ by construction.

A nice surprise, perhaps expected from the fact that $(1,1)\times \overline{(1,1)}$ might contain a $(2,2)$ factor, 
comes with ${\cal H}^{\rm long-}$: This combination of hypergeometrics 
has bosonic quantum numbers identified as  
\beq\label{Hminus_quantum_num}
\begin{array}{ccc}
1+\frac{\tau}{2}=\frac{\gamma}{2}+\kappa_2&\quad;\quad & l+1=\kappa_1-\kappa_2\ge 0   \\[.2cm]
\ \ \ \ \ \frac{b}{2}=\frac{\gamma}{2}-\kappa'_1 &\quad;\quad & a+1=\kappa'_1-\kappa'_2\ge 0
\end{array}
\eeq
where this time $\kappa'_1-\kappa'_2\ge 1$, by antisymmetry, and it is precisely the same  combination of hypergeometrics 
showing up in the long sector of $\mathcal{N}=4$ SYM \cite{Doobary:2015gia}. In the latter,
the Young diagrams are $(2,2)$ and come in differently, for example $\kappa_2=2+\frac{\tau-\gamma}{2}$ and $\kappa_1-\kappa_2=l$.
However, it is simple to see that the arguments of the$~_2F_1$, will coincide \cite{notagl22}. Thus the set of ${\cal H}^{\rm long-}$ 
is spanned by the same bosonic blocks that appear in $\mathcal{N}=4$ SYM in 4d.\\


{\bf Long two-particles operators with flavor.} The two-particle operators we want to study are long operators
exchanged in $\mathcal{M}^{\mathbb{S}}$ and $\mathcal{M}^\mathbb{A}$, and have the schematic form
\beq\label{two_part_def}
\mathcal{O}_{(rs)}^{+f}=\mathcal{P}^{+f}_{IJ} \Big[ \mathcal{O}^I_{r}\, \partial^l \Box^{\frac{1}{2}(\tau-r-s)}\mathcal{O}^J_s\Big]
\eeq
where $\mathcal{P}^{+f}$ is an appropriate projection for the flavor indexes. At leading order, 4pt diagrams in supergravity are those of a generalised disconnected free theory, and
by simple counting the two-particle operators above are degenerate. For given $SO(2,2)$ quantum numbers $\tau,l,$ and $SO(4)$ representation 
$\mathfrak{R}={[ab]}$, the number of degenerate states is nicely organised into a rectangle
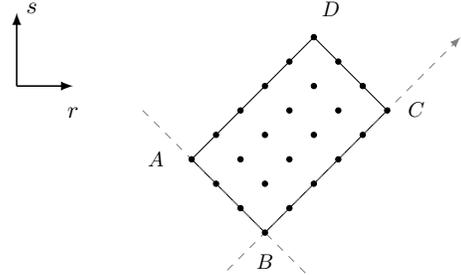
\beq\label{rettangolo_disegno}
\rule{.2cm}{0pt}
\begin{array}{c}
\begin{tikzpicture}[scale=.5]
%
%
\def\prop{.65}
\def\shifthor{\prop*2}
\def\ptuno{(\prop*2-\shifthor,\prop*8)}
\def\ptdue{(\prop*5-\shifthor,\prop*5)}
\def\pttree{(\prop*7-\shifthor,\prop*13)}
\def\ptquattro{(\prop*10-\shifthor,\prop*10)}
%
\draw[-latex, line width=.6pt]		(\prop*1   -\shifthor-4,         \prop*12          -0.5*\shifthor)    --  (\prop*1  -\shifthor-2.5  ,   \prop*12-      0.5*\shifthor) ;
\node[scale=.8] (oxxy) at 			(\prop*1   -\shifthor-2.5,  \prop*13.7     -0.5*\shifthor)  {};
\node[scale=.9] [below of=oxxy] {$r$};
%
\draw[-latex, line width=.6pt] 		(\prop*1   -\shifthor-4,     \prop*12       -0.5*\shifthor)     --  (\prop*1   -\shifthor-4,        \prop*15-      0.5*\shifthor);
\node[scale=.8] (oxyy) at 			(\prop*1   -\shifthor-1.8,   \prop*15.2   -0.5*\shifthor) {};
\node[scale=.9] [left of= oxyy] {$s$};
%
\draw[] 								\ptuno -- \ptdue;
\draw[black]							\ptuno --\pttree;
\draw[black]							\ptdue --\ptquattro;
\draw[]								\pttree--\ptquattro;
\draw[-latex,gray, dashed]					(\prop*0-\shifthor,\prop*10) --(\prop*8-\shifthor,\prop*2);
\draw[-latex,gray, dashed]					(\prop*3-\shifthor,\prop*3) --(\prop*13-\shifthor,\prop*13);
%
%
\foreach \indeyc in {0,1,2,3}
\foreach \indexc  in {2,...,7}
\filldraw   					 (\prop*\indexc+\prop*\indeyc-\shifthor, \prop*6+\prop*\indexc-\prop*\indeyc)   	circle (.07);
%
%
%
\node[scale=.8] (puntouno) at (\prop*4-1.5*\shifthor,\prop*8) {};
\node[scale=.8]  [left of=puntouno] {$A$};   
\node[scale=.8] (puntodue) at (\prop*5-\shifthor,\prop*5.5+.5) {};
\node[scale=.8] [below of=puntodue]  {$B$}; 
\node[scale=.8] (puntoquattro) at (\prop*11.2-\shifthor,\prop*12.5) {};
\node[scale=.8] [below of=puntoquattro] {$C$};
\node[scale=.8] (puntotre) at (\prop*7.7-\shifthor,\prop*11.7) {};
\node[scale=.8] [above of=puntotre] {$D$}; 
\node at  (8.5,\prop*13) {\phantom{space}};

\node [scale=.9] at (-2,0) {$
\begin{array}{l}
A=(a+1,a+b+1)\\[.1cm]
D=(a+t,a+b+t) \\[.1cm]
B=(a+\mu,a+b+2-\mu)\\[.1cm]
C=B+(t-1)
\end{array}$};

\node[scale=1] at (7,1.25) {$t= \frac{(\tau-b)}{2}-a$};
\node[scale=1] at (7,-0.5) {$\mu = \bigl\lfloor{\frac{b+1}{2}+\frac{1+(-1)^{a+l}}{4} }\bigr\rfloor$};

\end{tikzpicture}
\end{array}
\eeq
This rectangle $R_{\tau,l,[ab]}$ is analogous to the one in \cite{Aprile:2018efk}. 

Since the two-particle operators \eqref{two_part_def} are long, i.e.~non protected, they are expected to acquire an anomalous 
dimension of order $1/N$, i.e.~a binding energy in the gravity picture. We will compute their anomalous dimensions from the consistency of the OPE decomposition of the 4pt functions. 
We will only need two sets of matrix equations, given below in \eqref{equa_1} and \eqref{equa_2}, involving three-point couplings. 

Leading three-point couplings of the ${\cal O}^{+f}_{(rs)}$ with the external single-particle 
operators fit into a matrix
\beq
{\bf C}_{(pq), (rs)} \qquad (pq),(rs)\in R_{\tau,l,[ab]}
\eeq
where $(pq)$ refers to the pair of external operators ${\cal O}_p{\cal O}_q$, 
while $(rs)$ labels the two-particle operator.

The ${\bf C}_{(pq), (rs)}$ are found from disconnected Witten diagrams. 
These exist only for $\langle {\cal O}_p{\cal O}_q{\cal O}_p{\cal O}_q\rangle$. In the 
following we will  denote by $L({pqpq})$ the coefficients of their superconformal block 
decomposition  in the long sector. Note that because of the degeneracy, the three-point 
couplings are not one-to-one with  $L({pqpq})$, rather the data is organised in the form 
of matrix multiplication,
\beq\label{equa_1}
{\bf C}_{(p_1p_2),(rs)}\cdot {\bf C}^T_{(rs),(p_3p_4)}=\delta_{p_1p_3}\delta_{p_2p_4}\Big[ {\bf L}_{}^{}(p_1p_2p_3p_4)\Big]
 \eeq
as can be quickly derived from the OPE. There are ${\bf L}^{\pm}$ because we have two structure. Then, on a given $R_{\tau,l,[ab]}$ the matrices ${\bf L}^{\pm}$ are diagonal.

The anomalous dimensions of the ${\cal O}^+_{(rs)}$
enter the leading logarithmic discontinuity of the tree level correlators, through the equations 
\beq\label{equa_2}
{\bf C}_{(p_1p_2),(rs)}\cdot{\pmb \eta}^+\cdot{\bf C}^T_{(rs),(p_3p_4)}= {\bf M}(p_1p_2p_3p_4)
\eeq 
where ${\pmb \eta}^+$ is diagonal and ${\bf M}$ is defined from the decomposition
\beq\label{matrixM}
\!\!\!\!\begin{array}{l}
\displaystyle
		\mathcal{A}^+_{\vec{p}}\Big|_{\log U} 
		\!=\sum_{a,b}\ \frac{ {\tt B}^{(+p_{12},+p_{43})}_{-\frac{b}{2} ,\,-\frac{b}{2}-a}(y,\bar y) }{ \tilde U^{1-\frac{p_{4}+p_{3}}{2} } } \times \Bigg[  \\[.5cm]
\displaystyle
		\rule{2.2cm}{0pt}\sum_{\tau,l} {\bf M}_{\tau,l,a,b}(\vec{p}) \
		\frac{{\tt B}^{(-p_{12},-p_{43})}_{1+\frac{\tau}{2}+l,\, 1+\frac{\tau}{2} }(x,\bar x )}{ U^{1+\frac{ p_{4}+p_{3} }{2}  }} \Bigg]
\end{array}
\eeq

Some useful comments on this block decomposition are in order\footnote{Note that can assume without loss of generality that $\vec{p}$ is such that $p_{43}\ge p_{21}\ge 0$. This means for example that the diagram 
${\tt prefactor}_{\gamma=p_{4}-p_3}$ exchanges a dimension $={p_{43}}$ half-BPS operator, generalising the identity exchange for equal charges.} 
\begin{itemize}

\item For a given correlator $\tau\ge max(p_1+p_2,p_3+p_4)$.
It simple to see this inequality from Mellin space: Assume first 
$p_1+p_2\leq p_3+p_4$ then $c_s\leq 0$ and $s=0$ is the first double pole in \eqref{master_mellin}. Since the leading term from the r.h.s.~of \eqref{matrixM} goes like $U^{\frac{ \tau-{p_3+p_4} }{2}}$ we find $\tau=p_3+p_4$. Similarly, if $c_s\ge0$ then $s=c_s\ge 0$ is the first double pole, and by the same argument we now find the leading twist to be $\tau=c_s-p_3-p_4=p_1+p_2$.

\item The matrix ${\bf M}$ is full, i.e.~all entries are non trivial.  
\end{itemize}

At this point, normalising 
${\bf M}$ with the square root of ${\bf L}^+$ from the left and the right yields an \emph{unmixing matrix} whose 
eigenvalues are the anomalous dimensions, and the corresponding eigenvectors, the three-point 
couplings normalised. This is the same procedure adapted from  \cite{Aprile:2017xsp,Aprile:2018efk}, and we will give some explicit example in the next sections.
~ \\

{\bf Unmixing examples.} It is useful, before presenting general formulae, to exemplify the mixing problem in a few cases of interest. 
We will discuss first the symmetric flavor channel $f=\mathbb{S}$, and comment on $f=\mathbb{A}$ at the end, since the two will be related by a transformation.

The simplest rep we can study is $\mathfrak{R}={[00]}$. 
The first case we can look at is the unique two-particle operator at $\tau=2$ and even spin $l=0,2,\ldots 2\mathbb{N}$.  
This case has no mixing,
\beq\notag
\begin{array}{c}
{\bf L}_{\tau=2,l,[00]}= \frac{(l+1)!^2}{(2l+2)!}\times 2 \quad ;\quad
{\bf M}_{\tau=4,l,[00]}=\frac{(l+1)!^2}{(2l+2)!}\times 4
\end{array}
\eeq
are $1\times1$ matrices. 
The first mixing problem is at $\tau=4$, where we find two even spin operators. The corresponding data is
\beq
\begin{array}{rl}
{\bf L}_{\tau=4,l,[00]}=\frac{(l+2)!^2}{(2l+4)!}&\left[\begin{array}{cc}  \frac{2}{3} & 0 \\ 0 & \frac{1}{6}(l+1)(l+4)\end{array}\right] \\[.5cm] 
{\bf M}_{\tau=4,l,[00]}=\frac{(l+2)!^2}{(2l+4)!}&\left[\begin{array}{cc}  +4 & -4 \\ -4 & 10+5l +l^2\end{array}\right]
\end{array}
\eeq
Anomalous dimensions and three point functions are obtained by rewriting the mixing problem as an eigenvalue problem. 
In particular, for $\tau=4$ we find
\beq
{\bf L}^{-\frac{1}{2}}\cdot {\bf M}_{\tau=4,l,[00]} \cdot{\bf L}^{-\frac{1}{2}}={\bf c}\cdot\left[\begin{array}{cc}  -\tfrac{6(l+3)}{l+1} & 0 \\ 0 & -\tfrac{6(l+2)}{l+4}  0\end{array} \right]\cdot\,{\bf c}^T
\eeq
with
\beq\label{azero}
{\bf c}_{\tau=4,l,[00]}=\left[\begin{array}{cc} \sqrt{ \frac{l+1}{ 2l+5} } & +\sqrt{ \frac{l+4}{ 2l+5} }  \\ -\sqrt{ \frac{l+4}{ 2l+5} }  & \sqrt{ \frac{l+1}{ 2l+5} } \end{array}\right]
\eeq
an orthogonal matrix. 
The columns of this matrix ${\bf c}$ are the eigenvectors of the mixing problem. The leftmost eigenvector corresponds to the most negative anomalous dimension. 
We will label it with the leftmost corner of $R_{\tau,l,[00]}$. As the value of the anomalous dimension increases we move to the right of this corner.
The rep $[00]$ has no degeneracy thus $R_{\tau,l,[00]}$ is simply a line, and all anomalous dimensions are labelled uniquely. 

Next, let us consider the rep $\mathfrak{R}=[10]$. This is analogous to $[00]$, 
but for the fact that only odd spins contribute $l=1,3,\ldots,2\mathbb{N}+1$. The first two cases are $\tau=4$ with one operator,
\beq
\begin{array}{rl}
L_{\tau=4,l,[10]}= &\frac{(l+2)!^2}{(2l+4)!}\times\tfrac{1}{12} (24 + 25 l + 5 l^2) \\
M_{\tau=4,l,[00]}=&\frac{(l+2)!^2}{(2l+4)!}\times\, 2\,(l+1)(l+4)
\end{array}
\eeq
and $\tau=6$ with two operators. The mixing problem in this case is found from
\begin{align}
&
\!{\bf L}_{\tau=6,l,[10]}=\tfrac{(l+3)!^2}{(2l+6)!}\tfrac{(120 + 11(l +7)l)}{40} \left[\begin{array}{cc} 1& 0 \\ 0 & \frac{1}{9}(l+1)(l+6)\end{array}\right]\\
&
\!\!\begin{tikzpicture}
\draw (0,0) node[scale=.9] {$
{\bf M}_{\tau=6,l,[10]} =({\bf L}^{\frac{1}{2}}{\bf c})\!\cdot\!\!\left[\begin{array}{cc}  -\tfrac{120(l+4)(l+5)}{(120 + 11(l +7)l) } & 0 \\ 0 & -\tfrac{6(l+2)(l+3)}{(120 + 11(l +7)l)}  \end{array} \right]\!\!\cdot\!({\bf L}^{\frac{1}{2}}{\bf c})^T$};
\end{tikzpicture}\notag
\end{align}
where
\beq\label{anonzero}
{\bf c}_{\tau=6,l,[10]}=\left[\begin{array}{cc} \sqrt{ \frac{l+1}{ 2l+7} } & +\sqrt{ \frac{l+6}{ 2l+7} }  \\ -\sqrt{ \frac{l+6}{ 2l+7} }  & \sqrt{ \frac{l+1}{ 2l+7} } \end{array}\right]
\eeq
Comparing with the $2\times 2$ case from the $\mathfrak{R}={[00]}$, the free theory matrix ${\bf L}_{[10]}$ has some overall non factorisable term. By construction, this only affects the anomalous dimensions. 
The matrix ${\bf c}$ in \eqref{anonzero} has instead the same features as in \eqref{azero}. Quite remarkably, the matrix in \eqref{anonzero} is the same matrix that appears in the unmixing problem of [see \cite{Aprile:2017xsp}, eq.~(138)]  in $\mathcal{N}=4$ SYM in 4d.\footnote{To see this in formulae, we just need to realise that the values of the spins in \eqref{anonzero} are assumed to be odd, while those considered in  [\cite{Aprile:2017xsp}, eq.~(138)] were even.} 
We can help our intuition here by using Young diagrams.
In fact only Young diagrams that produce an $a\neq 0$ in $[ab]$ can be antisymmetrised in \eqref{change_b_block}, to yield the same as an 
$\mathcal{N}=4$ SYM block, as follows from the discussion below \eqref{Hminus_quantum_num}.
The case $a=1,b=0$ is the first case we find: We have originally two (1,1) Young diagrams $\underline{\kappa}_{i=1,2}$ with two rows, i.e.~$\underline{\kappa}_i=[\kappa_{i1},\kappa_{i2}]$, where 
\beq
\kappa_{i2}\in \{\varnothing,\Box\}\qquad;\qquad \kappa_{i1}\ge 1.
\eeq 
Therefore we find three $``+"$ blocks from 
$\varnothing\times \varnothing,\Box\times \Box,$ and $\varnothing \times \Box + \Box\times \varnothing$
and a single $``-"$ block from $\varnothing \times \Box - \Box\times \varnothing$.
As we discussed, the latter is the same as the $\mathcal{N}=4$ block that was used in  [\cite{Aprile:2017xsp}, eq.~(138)]. 

We close the list of examples by illustrating a mixing problem with partial degeneracy. The simplest case of partial degeneracy appears in
$\mathfrak{R}=[02]$, even spins $l=0,2,\ldots 2\mathbb{N}$ and $\tau=6$.\footnote{This rep also allows for odd spins, but there is no degeneracy for odd spins and we will not discuss it.} 
Note that $\tau=6$ is not the first available twist in the rep, which is instead $\tau=4$, but rather the next one.
In fact, $R_{\tau=6,l,[02]}$ consists of four points, and is a full rectangle, instead $R_{\tau=4,l,[02]}$ only covers a $-45^\circ$ oriented edge. 
The CFT data we are interested in to see the partial degeneracy is 
\beq
\!\!\!\!\!\!\!\!\!\!\!\!\!\!\!\!\begin{array}{l}
{\bf L}_{\tau=6,l,[02]}=\tfrac{(l+3)!(l+4)!}{(2l+6)!}\times  \\[.2cm]
\ \ \begin{tikzpicture}
\draw (0,0) node[scale=.9] {$
{\tt diag}\Big[\tfrac{2(l+4)}{15},\tfrac{3(l+1)(l+4)(l+6)}{160} ,\tfrac{3(l+3)}{5},\tfrac{(l+1)(l+3)(l+6)}{10} \Big]$};
\end{tikzpicture}
\end{array}\notag
\eeq
and
\beq
\!\!\!\!\begin{array}{c}
\begin{tikzpicture}
\draw (0,0) node[scale=.9] {$
{\bf M}_{\tau=6,l,[02]}=\frac{(l+3)!(l+4)!}{(2l+6)!}\times$};
\draw(2,-1.5) node[scale=.8] {$
\left[\begin{array}{cccc}  

				\frac{2(15+4L^2)}{3(-1+2L) } & \!\!-\frac{23+4L^2}{2(-1+2L) } & \ 8 & -8  \\[.2cm] 
				 -\frac{23+4L^2}{2(-1+2L) }  & \ \frac{ 1715 + 40 L^2+48 L^4}{128(-1+2L) } & \!-8 & \frac{55+4L^2}{8} \\[.2cm]
				 \ 8 & -8 & \frac{5+12 L^2}{1+2L} & -\frac{ 2(7+4L^2)}{ 1+2L } \\[.2cm]
				\! -8 & \frac{55+4L^2}{8}  & -\frac{ 2(7+4L^2)}{ 1+2L } & \frac{265-40L^2+16L^4}{8(1+2L)}
 \end{array}\right] $};
\end{tikzpicture}
\end{array}
\eeq
where we introduced $l=L-\frac{7}{2}$ just to simplify expressions. The eigenvalues are
\beq
\!\!\!\!\begin{array}{c}
\tfrac{1}{10} \pmb{\eta}_{\tau=6,l,[02]}=- {\tt diag}\Big[ \frac{(l+5)}{l+1}, \frac{(l+2)(l+5)}{(l+3)(l+4)},\frac{(l+2)(l+5)}{(l+3)(l+4)},\frac{(l+2)}{(l+6)}  \Big]
\end{array}
\eeq
The leftmost root is indexed, in $R_{\tau=6,l,[02]}$, by the leftmost corner at $(rs)=(13)$, then the two (degenerate) middle ones are indexed by $(rs)=(24),(22)$, and the rightmost by the rightmost corner at $(rs)=(33)$. 

Let us now comment that 
when we consider $f=\mathbb{A}$, what happens is that for given $\mathfrak{R}$ even and odd spin sectors are exchanged with respect to $f=\mathbb{S}$, but otherwise the mixing problem is the same. 
For this reason, we will restrict to $f=\mathbb{S}$ without loss of generality. ~\\

Our next task is to find general formulae for the mixing problem. We will begin with free disconnected theory, and then move to the anomalous dimensions.  
~\\


{\bf Cauchy identity for disconnected graphs.} The superconformal block decomposition 
of a free theory graphs, with $g_{ij}$ propagators connecting the various operators,  
can be done by decomposing the corresponding ${\cal C},{\cal S},\mathcal{H}^{\pm}$, 
for levels:  first ${\cal C}$, then ${\cal S}$ and finally $\mathcal{H}^{\pm}$, paying 
attention to include at each level the contributions from  the the previous ones. 
For a disconnected  Witten diagram, thus a disconnected graph, the first  
superconformal block contributing in each level has $\tau=p+q$, 
since  this equals the total number of bridges in the graph
going from ${\cal O}_{p} {\cal O}_{q}$ to ${\cal O}_{p} {\cal O}_{q}$, 
i.e. $\gamma=p+q$.

A more illuminating way of performing the same decomposition is
to use a Cauchy identity, as shown in \cite{Doobary:2015gia}. If $q>p$, 
there is one disconnected graph and the relevant identity is quite compact, 
\begin{align}\label{cauchy_id}
\!\!1\,&=\sum_{\ukap=[\kappa,1^{\kappa'-1}]}
\ A_{\gamma,\kappa+\frac{\gamma}{2}, \kappa'-\frac{\gamma}{2} }\ F^{(\alpha,\beta)}_{\gamma,\ukap}(x,y)\Big|_{\gamma=p+q} \\
\! A_{\gamma,\kappa,\kappa'}&=
\frac{ \Gamma[ \kappa\pm \frac{q-p}{2} ] \, \Gamma[2-2\kappa'] \, \Gamma[\kappa\pm\frac{\gamma}{2} ]^{\pm1} }{  
\Gamma[2\kappa-1]\, \Gamma[1- \kappa'\pm \frac{q-p}{2} ] \,\Gamma[ \frac{\gamma+1\mp1}{2} \pm \kappa']} 
\frac{(-)^{\kappa+\kappa'}}{\Delta^{(2)}_{\kappa,\kappa'} }\notag
\end{align}
with $\Delta_{\kappa \kappa'}^{(2)}=(\kappa'-\kappa)(\kappa+\kappa'-1)$. 
The mechanism behind \eqref{cauchy_id}  is quite neat: Recall that  $F_{\gamma,\underline{0}}=1+\ldots$ for $\beta\ge 1$ 
thus the r.h.s.~of \eqref{cauchy_id} is non trivial precisely because it has to compensate this half-BPS contribution.

From the  $(1,1)$ Cauchy identity, we obtain  the decomposition for the corresponding $AdS_3$ graph 
by taking $1\times \overline{1}$, expanding the sums and recognising,  out of the product, 
the relevant superconformal blocks. In the case $p=q$ there is  an additional graph contributing 
with $[(1-y)/(1-x)]^{\gamma/2}$. Since $p_{12}=p_{43}=0$,  this is related by crossing to 
\eqref{cauchy_id}, and we find again $(-)^{|\ukap|} A_{\gamma,\kappa,\kappa'}$. 
All together, the decomposition of disconnected graphs relative 
to $\mathcal{H}^{\rm long\pm}$, is 
\beq\label{Lplusminus}
\ \ \frac{ L^{\pm}_{\tau,l,[ab]}(pqpq)}{(1+\delta_{pq})}
		=\frac{ C_{\kappa,\kappa'}C_{\overline{\kappa},\overline{\kappa}'}\pm C_{\kappa,\overline{\kappa}'}C_{\overline{\kappa}, \kappa' }}{pq}
\eeq
where $C_{\kappa,\kappa'}=A_{p+q,\kappa+\frac{\gamma}{2},\kappa'-\frac{\gamma}{2}}$ ,
with the labels identified as in \eqref{change_qnumbers} for $L^+$ and as in \eqref{Hminus_quantum_num} for $L^-$. 
As we pointed out already, it should be the case that $L^-$ 
is the same as in $\mathcal{N}=4$ SYM. Indeed, 
\beq
L^{-}\!\propto {\tt Gammas} \times\! \frac{ (l+1)(a+1)(a+b+2)(\tau+l+2)}{\delta^{(8)} } 
\eeq 
where $\delta^{(8)}=\delta_{\kappa \kappa'}^{(2)} \delta_{\bar{\kappa} \bar{\kappa}'}^{(2)} \delta_{\kappa \bar{\kappa}'}^{(2)}\delta_{\bar{\kappa} \kappa'}^{(2)}$
and $\delta^{(2)}_{\kappa \kappa'}=\Delta^{(2)}_{\kappa+\frac{\gamma}{2},\kappa'-\frac{\gamma}{2}}$.
This in fact is the fully factorised formula obtained in \cite{Aprile:2018efk,Caron-Huot:2018kta,citaformula}, 
which we now discover to be secretly a $2$-by-$2$ determinant.\\
%


{\bf Anomalous dimensions.} Knowing the matrix ${\bf M}$, 
we can determine the anomalous dimensions.  Unfortunately it is hard to find a 
closed form expression for ${\bf M}$, but working out many cases we have found that 
the anomalous dimensions of ${\cal O}^{+f}_{(pq)}$ with $f=\mathbb{S}$
are consistent with the formula\footnote{The factor $\mathbb{D}$ introduced in \eqref{intro_anom_dim} is the part $\delta$ dependent of \eqref{new_anomalous_dim}.}
\begin{align}\label{new_anomalous_dim}
\eta^+_{\tau,l,[ab]}(rs)=&\ -\frac{2}{N}\frac{\delta^{(8)}}{
\delta_{\kappa \kappa'}^{(2)}\delta_{\bar{\kappa}\bar{\kappa}'}^{(2)}+\delta_{\kappa \bar{\kappa}'}^{(2)}
\delta_{\bar{\kappa}\kappa'}^{(2)}} \frac{1}{\rule{0pt}{.35cm}\left(\ell_{6d}+1\right)_2} \\[.2cm]
\ell_{6d}=&\ l+2r-a-1-\tfrac{1+(-1)^{a+l}}{2}
\end{align}
which is the main result of our paper. 

The $\eta^+_{\tau,l,[ab]}(rs)$ are simple rationals functions of the quantum numbers,
and very reminiscent  of the tree level anomalous dimensions for
$AdS_5\times S^5$ two-particle operators found in \cite{Aprile:2018efk}. Quoting from there
\begin{align}\label{quote_adS5}
\eta^{AdS_5\times S^5}_{\tau,l,[aba]}(rs)=&-\frac{2}{N^2}\ \delta^{(8)}\, \frac{1}{ (\ell_{10d}+1)_6} \\
\ell_{10d}=& l+2r-a-2-\tfrac{1+(-1)^{a+l}}{2}
\end{align}
Looking at \eqref{new_anomalous_dim}, we see now that  $AdS_5\times S^5$
tree level correlators were such that their decomposition in blocks
simplified $\delta_{\kappa \kappa'}^{(2)}\delta_{\bar{\kappa}\bar{\kappa}'}^{(2)}-\delta_{\kappa \bar{\kappa}'}^{(2)}
\delta_{\bar{\kappa}\kappa'}^{(2)}$ in the numerator of $L^-$. In fact, we also discover that this combination is fully factorised.  On the other hand, 
$\delta_{\kappa \kappa'}^{(2)}\delta_{\bar{\kappa}\bar{\kappa}'}^{(2)}+\delta_{\kappa \bar{\kappa}'}^{(2)}\delta_{\bar{\kappa}\kappa'}^{(2)}$ factorises only for $a=0$, otherwise it remains generic, 
as we saw in the unmixing examples in $\mathfrak{R}=[10]$. Thus only when $\mathfrak{R}=[0b]$ 
we find the simplification  $\delta^{(8)}/(\delta_{\kappa \kappa'}^{(2)}\delta_{\bar{\kappa}\bar{\kappa}'}^{(2)}+\delta_{\kappa \bar{\kappa}'}^{(2)}\delta_{\bar{\kappa}\kappa'}^{(2)})=\delta^{(4)}$ which is itself fully factorised. 

The partial degeneracy of the $\eta^+_{\tau,l,a,b}(rs)$ comes from the fact that they only depend on $r$, rather than $(rs)$, thus two-particle operators whose 
labels are on the same vertical axis in $R_{\tau,l,[ab]}$ have degenerate tree level anomalous dimension. This indeed is the same mechanism at work in \eqref{quote_adS5} for $AdS_5\times S^5$.
In comparison, the large spin behaviour goes like $-1/l^0$, and $-1/l^{2}$, in two, and four dimensions, respectively. 

Finally, for given $\mathfrak{R}=[ab]$, when $b$ is even, the set of anomalous dimensions is invariant under  $l\rightarrow -l-\tau-1$ in a given spin sector, 
and when $b$ is odd, the set of anomalous dimensions in the even spin sector is exchanged with those of the odd spin sector. This can be checked explicitly in the unmixing examples, and 
is just reciprocity symmetry. It was $l\rightarrow -l-\tau-3$ in $AdS_5\times S^5$.\\
%



{\bf Hidden symmetry.} We will now comment on the  tree level amplitudes for $AdS_{3}\times S^{3}$ (and $AdS_{5}\times S^{5}$)
discussed (and mentioned) above, from the point of view of a higher dimensional conformal symmetry. It will be convenient to introduce $\theta=2,4$ to parametrise $AdS_{\theta+1}\times S^{\theta+1}$,
then the tree level amplitudes for generic charges $\vec{p}$ descend from a single generating function, 
which is $\mathcal{A}_{\frac{\theta}{2}\frac{\theta}{2}\frac{\theta}{2}\frac{\theta}{2}}$
after a replacement of the cross ratios \cite{Caron-Huot:2018kta}. 
This specific $\mathcal{A}_{\frac{\theta}{2}\frac{\theta}{2}\frac{\theta}{2}\frac{\theta}{2}}$ 
is singlet under the sphere, and therefore the cross ratios of $SO(\theta,2)$ can be replaced 
with those of $SO(2\theta+2,2)$ canonically, e.g.~using the $AdS\times S$ Witten diagrams of \cite{Abl:2020dbx}. 
In Mellin space, this operation is the covariantisation $\mathcal{M}_{\vec{p}}=\mathcal{M}_{1111}({\bf s},{\bf t})$ for $AdS_{3}\times S^{3}$ and 
$\mathcal{M}_{\vec{p}}=\mathcal{M}_{2222}({\bf s},{\bf t})$ for $AdS_{5}\times S^{5}$ \cite{Aprile:2020luw}.

The parameter $\theta$ plays the role of dimensions in two ways. We have 
$\theta=d=2,4$ for the spacetime dimension of the CFT dual to the $AdS_{d+1}\times S^{d+1}$ gravity theory. 
Then we have $\theta=\frac{D-2}{2}$ where $D$ is the dimension of the flat background, i.e.~$D=2d+2=2\theta+2$, which is simply  $AdS_{d+1}\times S^{d+1}$ in conformally flat coordinates. 
This suggests that $\mathcal{A}_{\frac{\theta}{2}\frac{\theta}{2}\frac{\theta}{2}\frac{\theta}{2}}$ 
should then have a natural decomposition not only in long superconformal blocks for the corresponding SCFT, but also in $SO(2\theta+2,2)$ conformal blocks at the unitarity bound. 
Indeed, we find that
\beq\label{Simon1111}
\!\!\!\!\!\mathcal{A}_{\frac{\theta}{2}\frac{\theta}{2}\frac{\theta}{2}\frac{\theta}{2}}\Bigg|_{\log U}\!\!=\!
\sum_{\ell\ {} }\frac{ \theta\,\Gamma[\ell+\theta]^2 }{ \Gamma[2\ell+2\theta-1]} \frac{ ~_2F_1[ \theta+\ell,\theta+\ell;2\theta+2\ell; P] }{U^\theta}
\eeq
where$~_2F_1[ \theta+\ell,\ldots]$ is a single normalised block in which
we understand the$~_2F_1$ as a power series with the replacement 
$z^n\rightarrow P_{[ \theta+\ell+n,\theta]}(x,{\bar x};\theta)$, and $P(;\theta)$ being the two-variables Jack polynomial. A compact way of writing this polynomial is
\beq\label{Jack_pedro_notation}
\frac{(\theta)_{k} }{k!} \frac{ P_{[ \theta+k,\theta]}(x,\bar x;\theta)}{  U^{\theta} }= e^{-k\varphi}  \sum_{j=0}^{k } \frac{ (\theta)_j (\theta)_{k -j} }{ j! (k -j)! }\, e^{i(  k-2j)\phi}
\eeq
where $x=e^{-\varphi+i\phi}$ as in \cite{Bargheer:2019exp,Belitsky:2019fan} (and $k=\ell+n$). 

Quite remarkably the $SO(2\theta+2,2)$ decomposition in \eqref{Simon1111} only runs over a single sum. In fact, the second row of 
$P(;\theta)$ does not grow. To recover the usual double expansion 
over twist and spin of $SO(d,2)$ we need to 
recognise that within a $P_{\ukap}(;\theta)$ there are various $P_{\underline{\nu}}(;\theta')$ 
where $\theta'=(d-2)/2$ and $d=2,4$ \cite{Dolan:2003hv}, for $\theta=2,4$ respectively. In fact,
\beq\label{deco_jack}
\begin{array}{ll}
\displaystyle
\!\!\!P_{\ukap}(\theta)\!&=\,\sum_{m\ge 0} P_{[\kappa_1-m,\kappa_2+m]}(\theta')\times \frac{\ (-)^m}{m!} \times\bigg[\\[.3cm]
&\displaystyle
\ \ \ \ \frac{ (\theta-\theta'+1-m)_m\times (\kappa_-+1-2m)_{2m} }{(\theta+\kappa_- -m)_m (\theta'+\kappa_- -m)_m }\bigg]
\end{array}
\eeq
where $\kappa_-=\kappa_1-\kappa_2$, and $(\kappa_-+1-2m)_{2m}$ truncates the sum. 
Changing from $P_{\underline{\nu}}(;\theta')$ to the bosonic blocks \cite{Dolan:2003hv} gives the 
usual type of expansion. 

From the generating function, we obtain ${\cal A}_{\vec{p}}=\widehat{\cal D}^{}_{\vec{p}}\Big[ U^{\theta}\mathcal{A}_{\frac{\theta}{2}\frac{\theta}{2}\frac{\theta}{2}\frac{\theta}{2}} \Big]$, 
where
\beq
\widehat{\cal D}^{}_{\vec{p}}=\frac{ 1}{(U \tilde U)^{\frac{\,\theta}{2}}} \sum_{\tilde s,\tilde t} 
\left( \frac{\tilde U}{U}\right)^{\!\!{\tilde s}+\frac{\theta}{2}} \left(\frac{\tilde V}{V}\right)^{\!\!{\tilde t}} \widehat{\cal D}^{(0,0,0)}_{\vec{p},(\tilde s,\tilde t)} \widehat{\cal D}^{(c_s,c_t,c_u)}_{\vec{p},(\tilde s,\tilde t)}
\eeq
is a differential operator. 
As in \cite{Aprile:2020luw} we can find its explicit expression, 
\beq\label{operatorD}
\!\!\!\!\begin{array}{rl}
\widehat{\cal D}^{(a,b,c)}_{\vec{p},(\tilde s,\tilde t)} =&  {\displaystyle \frac{ (U\partial_U +1-\theta - \tilde s -a)_{\tilde s+a}}{(-)^a (\tilde s+a)!}} \times  \\[.3cm]
								\ \ \ \ \ \ \ &\times    {\displaystyle \frac{ (V\partial_V +1- \tilde t -b)_{\tilde t+b}}{(-)^b (\tilde t+b)!}}
								  {\displaystyle \frac{ (U\partial_U+V\partial_V)_{\tilde u+c}}{ (\tilde u+c)!}}
								 
\end{array}
\eeq
Understanding the action of  \eqref{operatorD} on $P(;\theta)$, for example in 
\eqref{Jack_pedro_notation},  might help finding an explicit formula for the three-point couplings.
Indeed, by acting with $\widehat{\cal D}^{}_{\vec{p}}$ on a single\,$~_2F_1[ \theta+\ell,\ldots;P]$ 
and summing, yields by construction  the mixing matrix, say on a ${R}_{\tau,l,[ab]}\otimes{R}_{\tau,l,[ab]}$ 
for reference.  As pointed out in \cite{Caron-Huot:2018kta}, this computation actually gives the 
mixing matrix as $\oplus_{r}(\eta\times\mathfrak{P}_r)$ where $\mathfrak{P}_r$ is a projector 
built out of the three-point couplings. In particular, these projectors descend from the$~_2F_1[ \theta+\ell,\ldots;P]$ 
and there are as many projectors as values of $r$ in $R_{\tau,l,[ab]}$. 
However, only when there is no residual degeneracy the projector is one-dimensional. 
%

Regarding the lift of the $AdS_3\times S^3$ partial degeneracy, let us comment that from the $AdS_{d+1}\times S^{d+1}$ Virasoro-Shapiro action 
postulated in \cite{Abl:2020dbx}, and specilised to our $d=2$ case, we have found evidence that 
a mechanism analogous to that discovered in \cite{Drummond:2020dwr,Aprile:2020mus}, 
will fix uniquely the three-point couplings.\\

{\bf Outlook.} 
The D1-D5 system has various tractable corners (see for example \cite{Maldacena:2000hw,Gaberdiel:2007vu,Pakman:2007hn,Eberhardt:2019ywk,Eden:2021xhe}), 
and most notably, the weak coupling regime has a (worldsheet) WZW description. 
But the 2d theory at the boundary of $AdS_3\times S^3$ with pure RR flux, 
whose 4pt correlators we have studied in this paper, is strongly coupled. 
The bootstrap approach is therefore quite natural in this case, since it does not rely on having a weakly coupled Lagrangian description. 
In fact, the many clues of hidden simplicity that we have encountered encourage the idea 
that our bootstrap program can tackle quantitatively this strongly coupled regime, offering new dynamical insights, beyond tree level. 
The clues we have found are neatly 
summarised by the form of the anomalous dimensions in \eqref{new_anomalous_dim}, and nicely 
accompanied by the structure of the generating function \eqref{Simon1111}, which provides the seed for the leading logarithmic discontinuity at any loop order. 
As in $AdS_5\times S^5$ we can now start computing one-loop correlators, and we will do so elsewhere.

On top of the above findings, we noted that the tree level dynamics (of tensor multiplets) on
$AdS_3\times S^3$ and that on $AdS_5\times S^5$ are aligned in many details. 
Worth mentioning is the fact that the chiral correlators, defined as \cite{Dolan:2003hv},
$\mathcal{G}^{\tt chiral}:={\cal G}\big|_{\bar y=\bar x}= \mathcal{C}+ (x-y) \mathcal{S}(x,y)$,
are actually equal in both theories, 
which suggests that the $(1,1)$ superconformal blocks `factorise' both theories, 
and tempts the idea that maybe there is a mechanism to understand the correlator, beyond the protected 
sector \cite{Bonetti:2016nma}, which still uses a Chern-Simons/WZW correspondence. 

More speculatively, it is also the case that when integrability techniques can be applied, some correlators \cite{Coronado:2018ypq} have a free fermion 
description \cite{Kostov:2019stn,Kostov:2021omc}, therefore it would be very interesting to understand 
whether the same is true for the $AdS\times S$ correlators we studied in this paper. 

Finally, the possibility of having $\mathcal{A}_{\vec{p}}^-\neq 0$ with $p_i>1$ remains open.\\

{\bf Acknowledgments.} We thank J.M.~Drummond, P.~Heslop and P.~Vieira for many 
discussions, and especially S.~Giusto, R.~Russo and the authors of \cite{nuovo_giusto} for providing 
important feedback and sharing their draft with us. MS thanks D.~Bufalini, H.~Paul and S.~Rawash 
for discussion.  FA is partially supported by the ERC-STG grant 637844- HBQFTNCER,
and MS by a Mayflower studentship from the University of Southampton.



\begin{thebibliography}{99}

\vspace{0.5cm}









\bibitem{Maldacena:1997re}
J.~M.~Maldacena,
Adv. Theor. Math. Phys. \textbf{2} (1998), 231-252
[arXiv:hep-th/9711200 [hep-th]].



\bibitem{deBoer:1998kjm}
J.~de Boer,
Nucl. Phys. B \textbf{548} (1999), 139-166
[arXiv:hep-th/9806104 [hep-th]].


\bibitem{Berkovits:1999im}
N.~Berkovits, C.~Vafa and E.~Witten,
JHEP \textbf{03} (1999), 018
[arXiv:hep-th/9902098 [hep-th]].
%

%
%
%
%
\bibitem{Taylor:2007hs}
M.~Taylor,
JHEP \textbf{06} (2008), 010
[arXiv:0709.1838 [hep-th]].





\bibitem{Romans:1986er}
L.~J.~Romans,
Nucl. Phys. B \textbf{276} (1986), 71


\bibitem{Deger:1998nm}
S.~Deger, A.~Kaya, E.~Sezgin and P.~Sundell,
Nucl. Phys. B \textbf{536} (1998), 110-140
[arXiv:hep-th/9804166 [hep-th]].


\bibitem{Mihailescu:1999cj}
M.~Mihailescu,
JHEP \textbf{02} (2000), 007
[arXiv:hep-th/9910111 [hep-th]].

\bibitem{Arutyunov:2000by}
G.~Arutyunov, A.~Pankiewicz and S.~Theisen,
Phys. Rev. D \textbf{63} (2001), 044024
[arXiv:hep-th/0007061 [hep-th]].









\bibitem{Penedones:2010ue}
J.~Penedones,
JHEP \textbf{03} (2011), 025
[arXiv:1011.1485 [hep-th]].
A.~L.~Fitzpatrick, J.~Kaplan, J.~Penedones, S.~Raju and B.~C.~van Rees,
JHEP \textbf{11} (2011), 095
[arXiv:1107.1499 [hep-th]].






\bibitem{Rastelli:2016nze}
L.~Rastelli and X.~Zhou,
Phys. Rev. Lett. \textbf{118} (2017) no.9, 091602
[arXiv:1608.06624 [hep-th]].
%
L.~Rastelli and X.~Zhou,
JHEP \textbf{04} (2018), 014
[arXiv:1710.05923 [hep-th]].



\bibitem{Alday:2017xua}
L.~F.~Alday and A.~Bissi,
Phys. Rev. Lett. \textbf{119} (2017) no.17, 171601
[arXiv:1706.02388 [hep-th]].



\bibitem{Aprile:2017bgs}
F.~Aprile, J.~M.~Drummond, P.~Heslop and H.~Paul,
JHEP \textbf{01} (2018), 035
[arXiv:1706.02822 [hep-th]].


\bibitem{Aprile:2017qoy}
F.~Aprile, J.~M.~Drummond, P.~Heslop and H.~Paul,
JHEP \textbf{05} (2018), 056
[arXiv:1711.03903 [hep-th]].




\bibitem{Alday:2018pdi}
L.~F.~Alday, A.~Bissi and E.~Perlmutter,
JHEP \textbf{06} (2019), 010
[arXiv:1809.10670 [hep-th]].



\bibitem{Drummond:2019hel}
J.~M.~Drummond and H.~Paul,
JHEP \textbf{03} (2021), 038
doi:10.1007/JHEP03(2021)038
[arXiv:1912.07632 [hep-th]].


\bibitem{Aprile:2019rep}
F.~Aprile, J.~Drummond, P.~Heslop and H.~Paul,
JHEP \textbf{03} (2020), 190
[arXiv:1912.01047 [hep-th]].


\bibitem{Bissi:2020woe}
A.~Bissi, G.~Fardelli and A.~Georgoudis,
[arXiv:2002.04604 [hep-th]].
%
A.~Bissi, G.~Fardelli and A.~Georgoudis,
[arXiv:2010.12557 [hep-th]].


\bibitem{Drummond:2020uni}
J.~M.~Drummond, R.~Glew and H.~Paul,
[arXiv:2008.01109 [hep-th]].








\bibitem{Aprile:2017xsp}
F.~Aprile, J.~M.~Drummond, P.~Heslop and H.~Paul,
JHEP \textbf{02} (2018), 133
[arXiv:1706.08456 [hep-th]].


\bibitem{Aprile:2018efk}
F.~Aprile, J.~Drummond, P.~Heslop and H.~Paul,
Phys. Rev. D \textbf{98} (2018) no.12, 126008
[arXiv:1802.06889 [hep-th]].

\bibitem{notagl22}
{See formulae (74) in section 3.1.1.~of \cite{Aprile:2017xsp}. }







\bibitem{Caron-Huot:2018kta}
S.~Caron-Huot and A.~K.~Trinh,
JHEP \textbf{01} (2019), 196
[arXiv:1809.09173 [hep-th]].



\bibitem{citaformula}
E.g.~formula (A.33) of \cite{Aprile:2020mus} has our same conventions.




\bibitem{Chester:2019pvm}
S.~M.~Chester,
JHEP \textbf{04} (2020), 193
[arXiv:1908.05247 [hep-th]].


\bibitem{Binder:2019jwn}
D.~J.~Binder, S.~M.~Chester, S.~S.~Pufu and Y.~Wang,
JHEP \textbf{12} (2019), 119
[arXiv:1902.06263 [hep-th]].






\bibitem{Giusto:2018ovt}
S.~Giusto, R.~Russo and C.~Wen,
JHEP \textbf{03} (2019), 096
[arXiv:1812.06479 [hep-th]].
%
%
\bibitem{Giusto:2019pxc}
S.~Giusto, R.~Russo, A.~Tyukov and C.~Wen,
JHEP \textbf{09} (2019), 030
[arXiv:1905.12314 [hep-th]].
%
\bibitem{Giusto:2020neo}
S.~Giusto, R.~Russo, A.~Tyukov and C.~Wen,
Eur. Phys. J. C \textbf{80} (2020) no.8, 736
[arXiv:2005.08560 [hep-th]].


\bibitem{Giusto:2020mup}
S.~Giusto, M.~R.~R.~Hughes and R.~Russo,
JHEP \textbf{11} (2020), 018
[arXiv:2007.12118 [hep-th]].


\bibitem{Rastelli:2019gtj}
L.~Rastelli, K.~Roumpedakis and X.~Zhou,
JHEP \textbf{10} (2019), 140
[arXiv:1905.11983 [hep-th]].








\bibitem{Aprile:2020luw}
F.~Aprile and P.~Vieira,
JHEP \textbf{12} (2020), 206
[arXiv:2007.09176 [hep-th]].


\bibitem{cross_ratios}
We use the same conventions as in \cite{Aprile:2020luw}



\bibitem{Lin:2015dsa}
Y.~H.~Lin, S.~H.~Shao, Y.~Wang and X.~Yin,
JHEP \textbf{12} (2015), 142
[arXiv:1508.07305 [hep-th]].


\bibitem{Heydeman:2018dje}
M.~Heydeman, J.~H.~Schwarz, C.~Wen and S.~Q.~Zhang,
Phys. Rev. Lett. \textbf{122} (2019) no.11, 111604
[arXiv:1812.06111 [hep-th]].






\bibitem{Dolan:2003hv}
F.~A.~Dolan and H.~Osborn,
Nucl. Phys. B \textbf{678} (2004), 491-507
[arXiv:hep-th/0309180 [hep-th]].
%
%
F.~A.~Dolan and H.~Osborn,
Annals Phys. \textbf{321} (2006), 581-626
[arXiv:hep-th/0412335 [hep-th]].

\bibitem{Doobary:2015gia}
R.~Doobary and P.~Heslop,
JHEP \textbf{12} (2015), 159
[arXiv:1508.03611 [hep-th]].







\bibitem{Giombi:2017cqn}
S.~Giombi, R.~Roiban and A.~A.~Tseytlin,
Nucl. Phys. B \textbf{922} (2017), 499-527
[arXiv:1706.00756 [hep-th]].




\bibitem{Cordova:2018uop}
L.~C\'ordova and P.~Vieira,
JHEP \textbf{12} (2018), 063
[arXiv:1805.11143 [hep-th]].



\bibitem{Drummond:2019odu}
J.~M.~Drummond, D.~Nandan, H.~Paul and K.~S.~Rigatos,
JHEP \textbf{12} (2019), 173
doi:10.1007/JHEP12(2019)173
[arXiv:1907.00992 [hep-th]].


\bibitem{Drummond:2020dwr}
J.~M.~Drummond, H.~Paul and M.~Santagata,
[arXiv:2004.07282 [hep-th]].~
%
\bibitem{Aprile:2020mus}
F.~Aprile, J.~M.~Drummond, H.~Paul and M.~Santagata,
[arXiv:2012.12092 [hep-th]].



\bibitem{Abl:2020dbx}
T.~Abl, P.~Heslop and A.~E.~Lipstein,
[arXiv:2012.12091 [hep-th]].






\bibitem{Bargheer:2019exp}
T.~Bargheer, F.~Coronado and P.~Vieira,
[arXiv:1909.04077 [hep-th]].

\bibitem{Belitsky:2019fan}
A.~V.~Belitsky and G.~P.~Korchemsky,
JHEP \textbf{05} (2020), 070
[arXiv:1907.13131 [hep-th]].








\bibitem{Maldacena:2000hw}
J.~M.~Maldacena and H.~Ooguri,
J. Math. Phys. \textbf{42} (2001), 2929-2960
[arXiv:hep-th/0001053 [hep-th]].
J.~M.~Maldacena and H.~Ooguri,
Phys. Rev. D \textbf{65} (2002), 106006
[arXiv:hep-th/0111180 [hep-th]].



\bibitem{Gaberdiel:2007vu}
M.~R.~Gaberdiel and I.~Kirsch,
JHEP \textbf{04} (2007), 050
[arXiv:hep-th/0703001 [hep-th]].
%
%
\bibitem{Pakman:2007hn}
A.~Pakman and A.~Sever,
Phys. Lett. B \textbf{652} (2007), 60-62
[arXiv:0704.3040 [hep-th]].


\bibitem{Eberhardt:2019ywk}
L.~Eberhardt, M.~R.~Gaberdiel and R.~Gopakumar,
JHEP \textbf{02} (2020), 136
[arXiv:1911.00378 [hep-th]].




\bibitem{Eden:2021xhe}
B.~Eden, D.~l.~Plat and A.~Sfondrini,
[arXiv:2102.08365 [hep-th]].






\bibitem{Bonetti:2016nma}
F.~Bonetti and L.~Rastelli,
JHEP \textbf{08} (2018), 098
[arXiv:1612.06514 [hep-th]].



\bibitem{Coronado:2018ypq}
F.~Coronado,
JHEP \textbf{01} (2019), 056
[arXiv:1811.00467 [hep-th]].
%
F.~Coronado,
Phys. Rev. Lett. \textbf{124} (2020) no.17, 171601
[arXiv:1811.03282 [hep-th]].


\bibitem{Kostov:2019stn}
I.~Kostov, V.~B.~Petkova and D.~Serban,
Phys. Rev. Lett. \textbf{122} (2019) no.23, 231601
[arXiv:1903.05038 [hep-th]].

\bibitem{Kostov:2021omc}
I.~Kostov and V.~B.~Petkova,
[arXiv:2102.05000 [hep-th]].


\bibitem{nuovo_giusto}
N.~Ceplak, S.~Giusto, M.~R.~R.~Hughes and R.~Russo,
[arXiv:2105.04670 [hep-th]].


%

 

\end{thebibliography}
\end{document}